\documentclass{aa}

\usepackage[dvips]{graphicx}
\usepackage{natbib}
\bibpunct{(}{)}{;}{a}{}{,}
\usepackage{amsmath}

\newcommand{\nIII}{N\,{\sc iii} $\lambda\lambda$4634,4642}
\newcommand{\nIV}{N\,{\sc iv} $\lambda$4058}
\newcommand{\nVa}{N\,{\sc v} $\lambda\lambda$4603,4619}

\newcommand{\heI}{He\,{\sc i} $\lambda$5875}

\newcommand{\heIIaaa}{He\,{\sc ii} $\lambda$4100}
\newcommand{\heIIa}{He\,{\sc ii} $\lambda$4200}
\newcommand{\heIIab}{He\,{\sc ii} $\lambda$4338}
\newcommand{\heIIb}{He\,{\sc ii} $\lambda$4542}

\newcommand{\heIId}{He\,{\sc ii} $\lambda$4859}

\newcommand{\cIV}{C\,{\sc iv} $\lambda$5808}

\begin{document}

\title{Another single hydrogen-rich Wolf-Rayet star in the SMC?\thanks{Based on observations made at the European Southern Observatory, La Silla, Chile.}}

\author{C. Foellmi\inst{1}}
\offprints{C. Foellmi, \email{cfoellmi@eso.org}}
\date{Received $<$date$>$ / Accepted $<$date$>$ }
\institute{European Southern Observatory, Alonso de Cordova 3107, Vitacura, casilla 19001,
Santiago, Chile}

\abstract{
A 12th Wolf-Rayet star in the SMC has recently been discovered by Massey et al. (2003). In order to determine its spectral type and a preliminary binary status, we obtained 3 high signal-to-noise spectra separated in time at the ESO-NTT. Compared to other WR stars in the SMC, SMC-WR12 appears to belong to the subgroup of faint, single and hydrogen-rich WN stars. We discuss the evolutionary status of WR12 and show that relatively low mass {\it rotating} progenitors can better account for the properties of single hydrogen-rich WN stars in the SMC.
}

\maketitle

\section{Introduction}

Until recently, the population of Wolf-Rayet (WR) stars in the SMC was considered nearly complete \citep[see the discussion in][]{Massey-Duffy-2001}. However, \citet*{Massey-etal-2003} have discovered a 12th WR star in the SMC from the same survey as \citet{Massey-Duffy-2001}. The discovery was delayed because of a misidentification of the target during the spectroscopic confirmation. According to their discovery data, it has an early spectral type in the nitrogen sequence (WN3-4.5) with a V magnitude of about 15.5. Unfortunately, these data do not allow a more precise determination of the evolutionary status of the star.

The discovery of another WR star in the SMC is important not only for the completeness issue, but also because the very low metallicity of the SMC makes it a very good laboratory to test the influence of metallicity on the evolution of WR stars, and the problem of their formation. This has a direct impact on studies of starbursts at low metallicity that use the SMC as a prototype \citep[e.g.][]{Schaerer-Vacca-1998}. 

Recently, the 10 previously known WN stars in the SMC have been studied by \citet*{Foellmi-etal-2003a}. These authors provided new and consistent spectral types, based on homogeneous high-S/N spectra. They have shown that a significant fraction, if not all WN stars have hydrogen absorption lines in their spectra and that these lines are clearly blue-shifted. They argued that hydrogen must be part of the WR, even for hot, early-type {\it single} WN stars. However, in their description of SMC-WR12, \citet{Massey-etal-2003} argue that "the presence of absorption spectra in the SMC WRs is still not well understood". 

In order to complete the study of \citet{Foellmi-etal-2003a} and to discuss the evolutionary status of this new interesting object, we have obtained a set of spectroscopic observations of SMC-WR12. This allows us to make a rapid check of radial-velocity (RVs) variations, to measure the RVs of the absorption lines present in the spectrum and to provide a reliable spectral type. In Section 2. we describe the observations and the RV measurements, while Section 3. contains the discussion. Section 4. summarizes our conclusions.

\section{Observations}

\begin{figure*}[!ht]
\centering
\includegraphics[width=\linewidth]{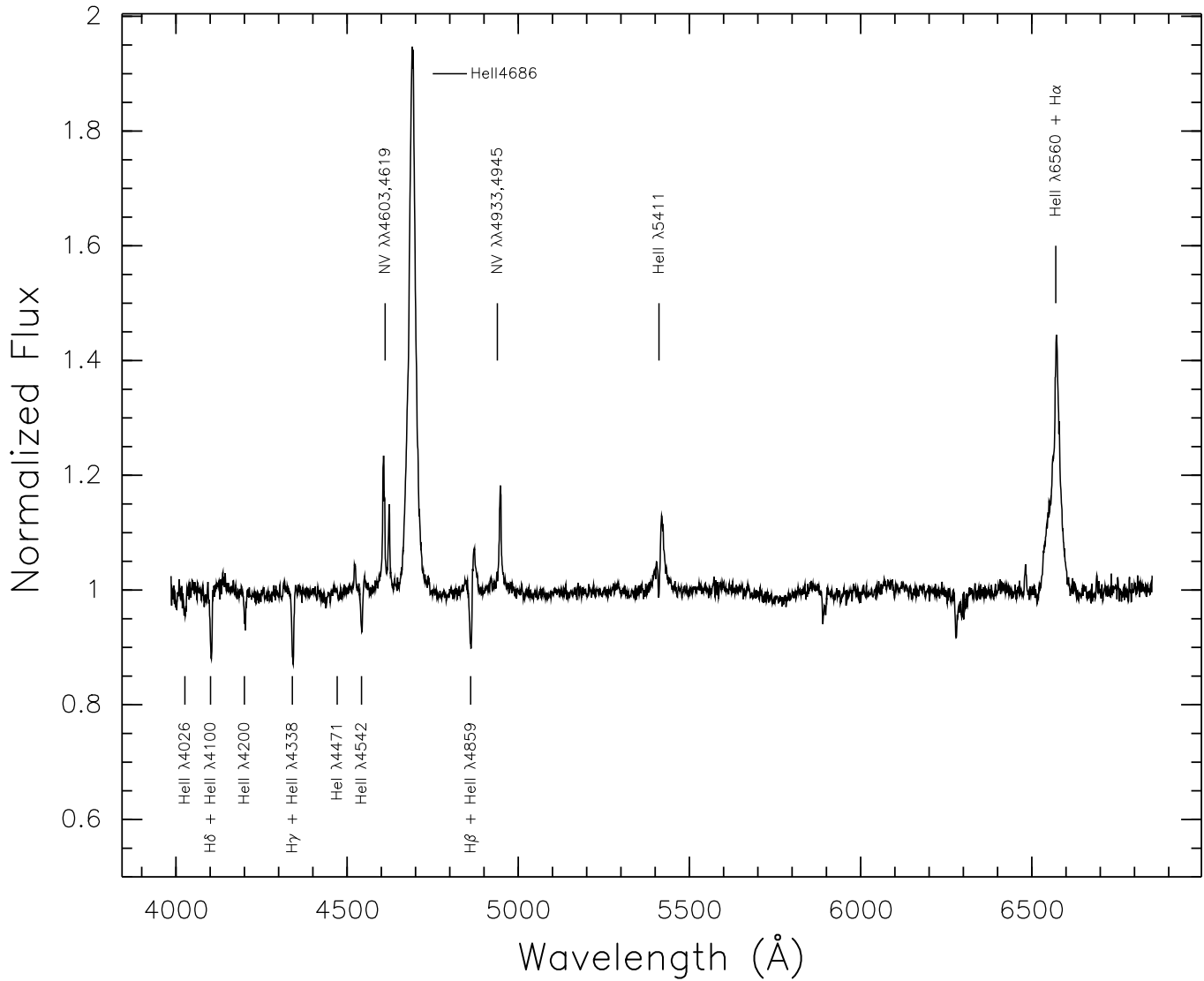}
\caption{Spectrum of SMC-WR12, obtained from a S/N-weighted mean of all spectra. We easily see hydrogen and helium absorption lines.}
\label{mean_spectrum}
\end{figure*}

We observed SMC-WR12 at the ESO-NTT during the nights of September 19, 20 and 29, 2003 (see the journal of observations in Table \ref{journal}). The reader is referred to the paper of \citet{Massey-etal-2003} for identification, magnitude, coordinates and finding chart. We used the spectrograph EMMI on the Red side, with Grism\#5 and a slit of 1''. This provides a resolution (FWHM) of $\sim$4.5 \AA. This wavelength range covers the most important emission \textit{and absorption} lines, from 4000 to 6800 \AA. In order to calibrate the spectra for radial-velocities (RVs), an arc spectrum was taken before and after the science exposures. Unfortunately, due to technical problems with the HeAr calibration lamp in the EMMI spectrograph during the third night, the only calibration spectrum we have for this night is less reliable. A rectified S/N-weighted mean spectrum of SMC-WR12 is shown in Fig.~\ref{mean_spectrum}.

To determine the spectral type we followed the classification scheme of \citet*{Smith-etal-1996}. In our spectrum, \heI\ is absent. Although not a defining criterium, this points towards a very early ionization subclass (WN2-3), as does the absence of \nIII, \nIV\ and \cIV. However \nVa\ is clearly visible, definitely making a WN3 ionization subclass for SMC-WR12. 

The time elapsed between the first and the last spectrum (10 days) allows us to perform a preliminary check for RV variations of WR12. For that purpose, we also observed a bright star (namely $\zeta$Dor = HD33262, spectral type F7V) with a known constant velocity to play the role of a RV standard. 

\begin{figure}
\includegraphics[width=\linewidth]{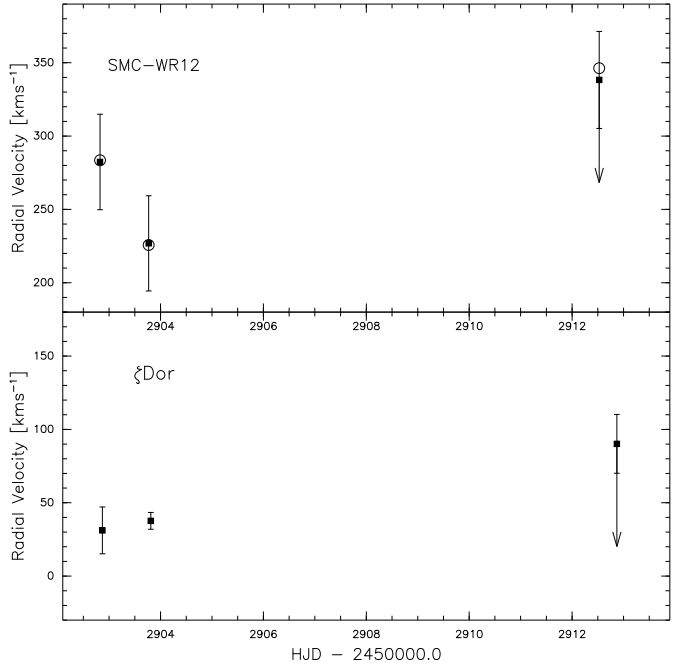}
\caption{Radial-velocities of SMC-WR12 (top), and $\zeta$Dor (bottom). Both panels have the same vertical scale to allow comparison. For SMC-WR12, the RVs have been measured with the cross-correlation method (filled squares) and the bissector method (open circles). For $\zeta$Dor, only cross-correlation has been used. The red-shifted calibration solution of the third night is easily visible for both stars (the arrow indicates -70 km\,s$^{-1}$). The RVs are consistant with a constant velocity (see text).}
\label{rvs}
\end{figure}

The spectra were first corrected to the heliocentric rest frame. Then, the RVs of the 2 stars were measured with a cross-correlation (CC) technique and bissectors. The procedure used is identical to the one described in \citet{Foellmi-etal-2003a}, and the wavelength window for CC ranges from 4570 to 4800 \AA\ to avoid any possible influence from the absorption lines. The RVs of SMC-WR12 and $\zeta$Dor are plotted in Fig.~\ref{rvs}. The figure shows that the spectrum of the third night is red-shifted for both stars, by a similar amount ($\sim$70 km\,s$^{-1}$).

\begin{table}
\centering
\caption{Journal of observations. The heliocentric Julian Date, the signal-to-noise ratio in the continuum and the total exposure time (in minutes) are indicated.}
\label{journal}
\begin{tabular}{lrc} \hline
HJD		& S/N 	& total exp. \\ \hline
2452902.8243	&  86	&  85 \\
2452903.7696	&  99 	&  60 \\
2452912.5273 	& 121	& 120 \\ \hline
\end{tabular}
\end{table}

Although we cannot rule out the possibility, with only 3 spectra over 10 days, that SMC-WR12 could be a binary star with a low-inclination angle, the RVs presented here are consistent with a single star (after correction of the third night). Additional data will obviously be needed to confirm this preliminary result.

To add weight to this presumed single star status, we have measured the RVs of the absorption lines, using a gaussian profile, on the (heliocentric) spectrum of each night. The central wavelengths of the lines were compared to their rest wavelength. The results are summarized in Table~\ref{rv_abs}. 

Similar to what is seen in other single WN stars in the SMC, these lines have a constant RV and are clearly blue-shifted relative to the frame of the star, whose mean measured RV (from the CC, see above) is $\sim$260 km\,s$^{-1}$. However, the emission lines of WR stars are known to not reflect the true systemic velocity of the star, since the line profiles are often not symmetric, and might be distorted by the presence of an even weak P~Cygni profile. Nevertheless, when compared to the mean RV of the SMC (160 km\,s$^{-1}$), the blue shift observed in the RVs of the absorption lines of WR12 remains significant. 

Note that for the blended lines, the resolution of the spectra does not allow a precise fitting by a double gaussian (the resolution is $\sim$ 300 km\,s$^{-1}$ while the lines are separated by $\sim$ 123 km\,s$^{-1}$). The measured central wavelength has been compared to the simple mean of the rest wavelengths of the two lines. If the hydrogen line is dominant in the blend, the radial velocity measurement quoted in Table~\ref{rv_abs} is an upper limit. This makes even stronger the blueshift. If, on ther hand, the helium line is dominant, a shift of 123/2=62 km\,s$^{-1}$ needs to be added, and this would make the blueshift less obvious. However, the hydrogen influence must be nonetheless significant since blended lines have larger equivalent widths.

Therefore, the blueshift is real and the absorption lines of SMC-WR12 must be formed deep in the relatively weak WR wind, at modestly blue-shifted RV along the line of sight to the stellar core. They do belong to the WR object itself, and cannot come from a binary companion (that would also contradict the combined observed constancy of the RVs). 

Moreover, as argued in \citet{Foellmi-etal-2003a}, the SMC WR stars whose emission lines do not satisfy the broad-line criterium in the classification scheme of \citet{Smith-etal-1996} contain hydrogen \citep[see][]{Smith-Maeder-1998}. Additionaly, the presence of hydrogen can be assessed from the Balmer versus Pickering decrement. As seen in Table~\ref{rv_abs}, the equivalent widths of the blended H+He lines are larger than that of pure He lines, which indicates the presence of hydrogen in the spectrum \citep[see also][ for a detailed discussion]{Hamann-etal-1991}. Although being defined for emission lines, the second criterium for the hydrogen in the Smith et al. classification (see their Table~4c) can be computed since the three relevant lines are all in absorption. The value obtained is 1.9 for SMC-WR12, making a strong case in favour of an "h" suffix.

\begin{table}
\centering
\caption{Radial-velocities and equivalent widths of absorption lines of SMC-WR12. The rest wavelength used for each (combination of) lines \citep{wavetable} is also indicated. For the combination of 2 lines, the mean wavelength has been computed. The RV is the mean of RVs measured on the 3 spectra. The estimated accuracy is $\sim$30 km\,s$^{-1}$ on individual measurement. The accuracy has been estimated by making the gaussian fit with different fraction of the line and/or the continuum. The variance of these various results give an estimation of the accuracy. Moreover, the accuracy can easily reach a fraction of pixel, usually a tenth of the resolution. With a 300 km\,s$^{-1}$ resolution, the accuracy is about 30 km\,s$^{-1}$.}
\label{rv_abs}
\begin{tabular}{lrrr} \hline
Line & $\lambda_{0}$ (\AA) & RV (km\,s$^{-1}$)& EW (\AA) \\ \hline
H$\delta$ $\lambda$4101 + \heIIaaa	& 4100.87 & 58 & 0.76 \\
\heIIa								& 4199.85 & 35 & 0.40 \\
H$\gamma$ $\lambda$4340 + \heIIab	& 4339.58 & 53 & 0.94 \\
\heIIb								& 4541.66 & -9 & 0.57 \\
H$\beta$ $\lambda$4861 + \heIId		& 4860.33 & -34& 1.17 \\ \hline
\end{tabular}
\end{table}

Finally, the complete spectral type of WR12 is WN3ha. This spectral type makes even stronger the need for a complementary classification, as proposed by \citet{Foellmi-etal-2003b}, since the star has a very early ionization subclass and a significant hydrogen content at the same time, the latter being more related to late-type WN stars. In that context, the evolutionary classification of SMC-WR12 is eWNL.

\section{Discussion}

This 12th WR star in the SMC, if proved to be single as seems to be the case, would lower even more the binary frequency of the SMC ($\sim$40\%) found by \citet{Foellmi-etal-2003a}. However, with only 12 stars, it remains small number statistics. Nevertheless, the consequence of the existence of this new member is that nearly half of the WR population is hydrogen-rich with an early spectral type.

\subsection{Retort to the conclusions of Massey et al.}

\citet{Massey-etal-2003} argue that the origin of absorption lines in WR spectra is either due to the fact that the WR winds are weak and we see photospheric lines, or that these stars are binaries. Both suppositions are in contradiction with the fact that these lines are strongly blue-shifted. 

If these absorption lines come from the presupposed "photosphere" of the WR star, it would be impossible to reach values as low as 50 km\,s$^{-1}$ when the mean velocity of WR12 is about 160 km\,s$^{-1}$ (assuming that the star has a systemic velocity similar to that of the SMC). This velocity difference is too large for a photosphere, and these lines must be formed in the wind itself.

As for binaries, they admit that the study by \citet{Foellmi-etal-2003a} "suggest that the binary fraction of WRs in the SMC is normal". This is half the truth. Foellmi et al. discussed the binary status of each of the WN stars in the SMC with not only the help of radial-velocities, but also line-profile variations, absolute magnitudes, spectral classification from a homogeneous set of spectra, archival photometric data and X-rays fluxes for most of the stars. Although the case of SMC-WR9 was not definitive, they have shown that SMC-WR1, WR2, WR4, WR10 and WR11 are certainly single stars, and they do show absorption lines in their spectra. WR12 is very similar to these WN stars in the SMC, especially WR1 and WR10.

We think that the presence of intrinsic hydrogen in the spectra of single WN stars in the SMC is inescapable. The H/He ratio (by number), obtained from a preliminary modelling of the spectra (Crowther 2002, private communication) even reaches values as high as 1 for WR9 and 2 for WR10. The spectral types of all the single WN stars in the SMC are reproduced in Table~\ref{types}, along with the absolute magnitudes. It can be seen that WR12 fits perfectly well in this subgroup of stars, especially with the faintest.

\begin{table}
\centering
\caption{Spectral types and absolute magnitudes of single WN stars in the SMC. The magnitudes are from \citet{Massey-etal-2003}, although that of WR11 has been corrected as discussed in \citet{Foellmi-etal-2003a}. WR12 appears as the hottest, and the second faintest WN star with hydrogen.}
\label{types}
\begin{tabular}{lcc} \hline
Star		& Spectral Type	& M$_{V}$	\\ \hline
WR1			& WN3ha 		& -4.6	\\
WR2			& WN5ha			& -5.2	\\
WR4			& WN6h			& -6.2	\\
WR9			& WN3ha			& -4.3	\\
WR10		& WN3ha			& -3.6	\\ 
WR11		& WN4h:a 		& -4.7	\\
WR12		& WN3ha 		& -4.0	\\ \hline
\end{tabular}
\end{table}

The presence of hydrogen has important implications on the formation of WR stars at low metallicity, and more precisely on the initial mass of their progenitors.

\subsection{The problem of the initial mass}

As in \citet*{Massey-etal-2000}, \citet{Massey-etal-2003} claimed that WR stars in the SMC come from only the highest mass (greater than 70$M_{\odot}$) stars, in accord with the expectations that at low metallicities only the most massive and luminous stars will have sufficient mass loss to become WR stars. 

There is, in our sense, one important caveat in \citet{Massey-etal-2000} that led to wrong conclusions. They do not account for the influence of rotation \citep[e.g. using the "old" evolutionary paths of][]{Schaller-etal-1992}, although numerous pieces of evidence exist today that rotational effects modify strongly the evolution of massive stars, and therefore the shaping of a WR population at low metallicity. More precisely, rotating stars appear overluminous for their actual masses \citep{Maeder-Meynet-2000}. Consequently, the estimation of the progenitor masses of WR stars in the Magellanic Clouds from cluster turn-offs are certainly too high, in addition to being in contradiction with the presence of hydrogen in faint single WN stars.

The problem with very high initial mass progenitors at low $Z$ is that they cannot explain the concomitant facts that WN stars in the SMC have a very early ionization subclass and hydrogen in their wind. These stars cannot have had high initial mass progenitors, since such masses would mean a high mass-loss during the Main-Sequence phase, and the hydrogen-rich envelope would have been removed by the time the WN phase was reached. If, because of the metallicity, the mass-loss is drastically reduced and some hydrogen is still left in the envelope, it is incompatible with the early ionization subclasses observed, because more massive WN stars are cooler, leading to a late spectral type.

On the other hand, relatively "low" mass (say 30-50$M_{\odot}$) {\it rotating} progenitors are believed to produce early hydrogen-rich WR stars at low metallicity.

\subsection{The question of rotation}

Rotation favours the formation of WR stars by two means: an increased internal mixing and an enhanced mass-loss rate during the MS phase \citep{Maeder-Meynet-2001}. In the SMC, a mass-loss rate enhanced by rotation would be balanced by the low metallicity, while the enhanced internal mixing will drive hydrogen down to deeper layers, leading to WR stars that will remain "hydrogen-rich" longer. We can therefore expect theoretically similar values of the minimum mass of formation of a WR star, whatever the metallicity.

Moreover, in the context of the absence of rotation, the WR binary frequency was expected to be clearly metallicity-dependent, since fewer single WR stars will be formed compared to those formed in binaries.\footnote{We make the hypothesis that the influence of a companion on the formation of the WR component is independent of the metallicity.} However, the binary frequency among WR stars has been shown to be similar, and statistically identical, both in the SMC \citep{Foellmi-etal-2003a} and the LMC \citep*{Bartzakos-etal-2001,Foellmi-etal-2003b}, and compatible with the galactic frequency, although this latter value is not reliable because of incompleteness. In fact, these similar values hide the opposite contribution of two different effects: at low Z, the mass-loss rate is lower, thus less angular momentum is lost during the Main Sequence, and the WR progenitors will have a larger mean rotation velocity \citep[see e.g.][]{Maeder-Meynet-2000}, favouring the formation of WR stars even for relatively low mass progenitors.

As for the early types observed in the SMC, \citet{Crowther-2000} has shown that for identical physical parameters (except the metallicity), wind modelling gives earlier subtypes at lower metallicity. As a matter of fact, if the rotation enhances the internal mixing, the increased abundance of helium in the atmosphere of the star reduces its opacity. The star will be more compact, bluer, and for a given mass it will be hotter. Since the temperature is directly related to the ionization subclass \citep*[e.g.][]{Hamann-etal-1995}, it explains the early types observed.

While the influence of rotation in short-period binaries is still hard to evaluate, it seems clear that at least the single WR stars in the SMC have been formed, or at least strongly influenced, by rotation.

\section{Conclusion}

We have obtained a set of 3 high-S/N spectra spread over 10 days of the recently discovered WR star in the SMC. We have measured its RVs and found them to be consistent with a constant velocity, indicating a probable single-star status. We have also measured the RVs of hydrogen and helium absorption lines found in the spectra. Similar to what is seen in other SMC WN stars, the absorption lines have constant radial-velocities and are strongly blue-shifted. Consequently, hydrogen must be present in the wind of the star, and the lines are formed along the line of sight through the relatively weak wind. Our spectral classification is WN3ha, while the evolutionary classification is eWNL.

The presence of hydrogen in single WN stars in the SMC raised the question of the mass of their progenitors, and the influence of metallicity and rotation on the formation of WR stars at low Z. We have shown that rotation effects on relatively "low" mass progenitors can better explain the whole set of properties of SMC-WR12 and other single, faint, hydrogen-rich WN stars in the SMC.

\begin{acknowledgements}
C.F. is grateful to K.Aubel and O.R.Hainaut. C.F. also thanks L.Germany for a careful reading of the manuscript, and A.F.J.Moffat for valuable comments and for having brought to his attention the preprint of Massey et al. The author also acknowledges the referee (W.-R. Hamann) for comments and important improvements on the manuscript. 
\end{acknowledgements}

\bibliographystyle{aa}
\bibliography{mnemonic,wr12bib}

\end{document}